\documentclass[amssymb,floatfix,aps,twocolumn]{revtex4}
\emergencystretch=\maxdimen
\usepackage{euler}
\usepackage{graphicx}
\usepackage{dcolumn}
\usepackage{amsmath}
\usepackage[latin1]{inputenc}
\usepackage{amssymb}
\usepackage[colorlinks=true, citecolor=blue, urlcolor = blue, linkcolor= red, bookmarks=true]{hyperref}
\usepackage{float}
\usepackage{amsfonts}
\usepackage{dcolumn}
\usepackage{hyperref}
\usepackage{subfigure}
\usepackage{pgfplots}
\usepackage{booktabs}
\usepackage{mathrsfs}
\usepackage{multirow}
\pgfplotsset{compat=1.16}

\usepackage{wrapfig}
\usepackage{wrapfig}
\usepackage{adjustbox}
\makeatletter
\def\btt#1{\texttt{\@backslashchar#1}}
\DeclareRobustCommand\bblash{\btt{\@backslashchar}} \makeatother

\begin{document}

\title[]{Ehrenfest Paradox: A Careful Examination}
\author{Jitendra Kumar} \email{jitendra0158@gmail.com}

\affiliation{Centre for Theoretical Physics, Jamia Millia Islamia, New Delhi 110025, India}
\begin{abstract}
The Ehrenfest paradox for a rotating ring is examined and a kinematic resolution, within the framework of the special theory of relativity, is presented. Two different ways by which a ring can be brought from rest to rotational motion, whether by keeping the rest lengths of the blocks constituting the ring constant or by keeping their lengths in the inertial frame constant, are explored and their effect on the length of the material ring in the inertial as well as the co-rotating frame is checked. It is found that the ring tears at a point in the former case and remains intact in the latter case, but in neither of the two cases is the motion of the ring Born rigid during the transition from rest to rotational motion.

\end{abstract}

\maketitle
\thispagestyle{empty}

\setcounter{page}{1}

\section{Introduction}\label{sec1}
In 1909, after Max Born introduced a notion of rigid motion \cite{Born} in special relativity, Paul Ehrenfest presented a paradox about a cylinder that goes from rest to rotational motion which gives contradicting results if the notion of Born rigidity is applied to the cylinder. Born rigidity or equivalently Born rigid motion is defined as: ``Let there be a continuum of infinitesimal observers who travel along with the points of the non-uniformly moving body: for each of them in their measure the infinitesimal neighborhood should appear permanently undeformed" \cite{Born, Ehrenfest}. It was shown in Refs. \cite{Born, Herglotz} that the definition of Born rigidity has nothing to do with the property of the material, how flexible or rigid that material is; rather it is about a special kind of motion of an extended object that preserves the distance between the nearby points of the object for the observers traveling along with the points. A perfectly rigid body cannot exist in nature as it will violate relativity but it is possible, in some instances, to move a body in Born rigid way as shown in Refs. \cite{Herglotz, Noether}. This definition of Born rigid motion is equivalent to Ehrenfest's relative-rigid motion \cite{Ehrenfest}. The paradox arises when this definition of Born rigidity is applied to a cylinder going from rest to rotational motion. 

Stepanov \cite{Stepanov:2013sia,Stepanov2018} formulated three different definitions of rigidity of a non-inertial frame of reference in motion with respect to an inertial frame:
\begin{enumerate}
\item Comoving rigidity: all points of the reference frame have zero velocity in the inertial frame co-moving relative to one of its points.
\item Local rigidity: the tensor $\gamma_{ij} = -g_{ij} + g_{0i} g_{0j}/g_{00}$ defining the element of infinitesimal physical length does not depend on time.
\item Global rigidity: the radar distance (distance measured by an observer who sends out a light signal from one point to another and then receives it back) between any two points of the reference frame is constant.
\end{enumerate}
Stepanov \cite{Stepanov:2013sia,Stepanov2018} showed that the above three definitions are not mutually equivalent and also demonstrated that Born's definition of rigidity, if cast in covariant notation, coincides with the local rigidity. As the Ehrenfest paradox concerns itself with the Born rigidity, we will mainly focus on Born rigidity.

The paradox, in Ehrenfest's words, but with symbols chosen to be consistent with this paper's presentation, is stated as \cite{Ehrenfest}:
``Let a relative-rigid cylinder of radius $r$ and height $h$ be given. A rotation about its axis which is finally constant, will gradually be given to it. Let $r_{1}$ be its radius during this motion for a stationary observer. Then $r_{1}$ must satisfy two contradictory conditions: 

a) The periphery of the cylinder has to show a contraction compared to its state of rest: $2\pi r_{1}<2\pi r$, because each element of the periphery is moving in its own direction with instantaneous velocity $r_{1}\omega$.

b) Taking any element of a radius, then its instantaneous velocity is normal to its extension; thus the elements of a radius cannot show a contraction compared to the state of rest. It should be: $r_{1}=r$."

And hence, it is concluded that a cylinder cannot be brought from rest to rotational motion while maintaining the notion of Born rigidity. But it does not answer in what way the cylinder has to be deformed during the transition from rest to rotation. If the periphery contracts while the radius remains the same, this seems to imply that it must shatter at the periphery. However, this raises another question: What are the locations where it should shatter and why are those locations preferred over others, as a cylinder is perfectly symmetric about its axis of rotation.

Many attempts \cite{Sama, Ives, Cavalleri, Cantoni, Gron1975, Weber} have been made to analyze the paradox especially for the rotating disk but there is no general consensus in literature on the solution of the problem. Sama \cite{Sama} proposed that the paradox arises ``from an ambiguous use of notation" and is not actually connected with any inconsistency in relativity. Ives \cite{Ives} expressed the view that the plane of the rotating disk bends taking the shape of a ``dish''. But this kind of deformation, as pointed out by Cavalleri \cite{Cavalleri}, is not symmetric with respect to the plane of the non-rotating disk and thus violates spatial parity. Planck \cite{Planck} argued that the resolution will require employing the theory of elasticity as a rigid body does not exist in nature. Pursuing a similar line of thought, Cavalleri \cite{Cavalleri} presented a dynamical resolution of the paradox and argued that no kinematic resolution of the paradox is possible. The argument by Cavalleri was refuted by Cantoni \cite{Cantoni} who claimed to show, on purely kinematical grounds, ``that one of the assumptions implicitly contained in the statement of Ehrenfest's paradox is not correct, the assumption being that the geometry of Minkowski space-time allows the passage of the disk from rest to rotation in such a fashion that both the length of the radius and
the length of the periphery, measured with respect to the comoving frame of reference, remain unchanged.'' Gr{\o}n \cite{Gron1975, Gron2004} presented a kinematic resolution of the paradox which, according to Gr{\o}n, stems from the impossibility of accelerating the infinitesimally close points at the periphery of the disk simultaneously in the successive rest frames. Weber \cite{Weber} claimed that the material of the disk will physically stretch in the circumferential direction, thus increasing the rest length of the periphery which compensates for its Lorentz contraction during the transition from rest to rotational motion. Ghosal \textit{et al.} \cite{Ghosal} argued that, for Weber's claim to hold, there is an implicit assumption that the length of the periphery in the inertial frame remains the same during the transition from rest to rotation. Rizzi and Ruggiero \cite{Rizzi:2002sk, Ruggiero} analyzed the spatial geometry of a rotating disk, which they found to be non-Euclidean and using non-Euclidean geometry they claimed to have given a resolution of the paradox. But it needs to be pointed out that the paradox is about the transition of a disk from rest to rotational motion and not about a disk in steady-state rotational motion with constant angular velocity. So, analysis of the spatial geometry of a rotating disk, while being non-Euclidean, does not resolve the actual paradox posed by Ehrenfest. Also, the notion of Born rigidity is maintained for a disk in steady-state rotational motion, the problem with Born rigidity arises during the transition from rest to rotation \cite{Gron1979}.

In this article, the Ehrenfest paradox for a ring is analyzed and a resolution, in the domain of the special theory of relativity, is provided. A distinction is made between two scenarios, a ring in steady-state rotational motion and the transition of a ring from rest to rotation, and it is explained that a paradox arises only in the latter scenario. Two natural ways of accelerating a rod, keeping its length in the inertial frame constant or keeping its proper length constant, are explained, depending on which two different ways by which a ring can be brought from rest to rotational motion are explored and it is examined what effect it has on the length of the material ring as measured in the inertial or co-rotating frame.

\begin{figure} 
		   \includegraphics[scale=0.7]{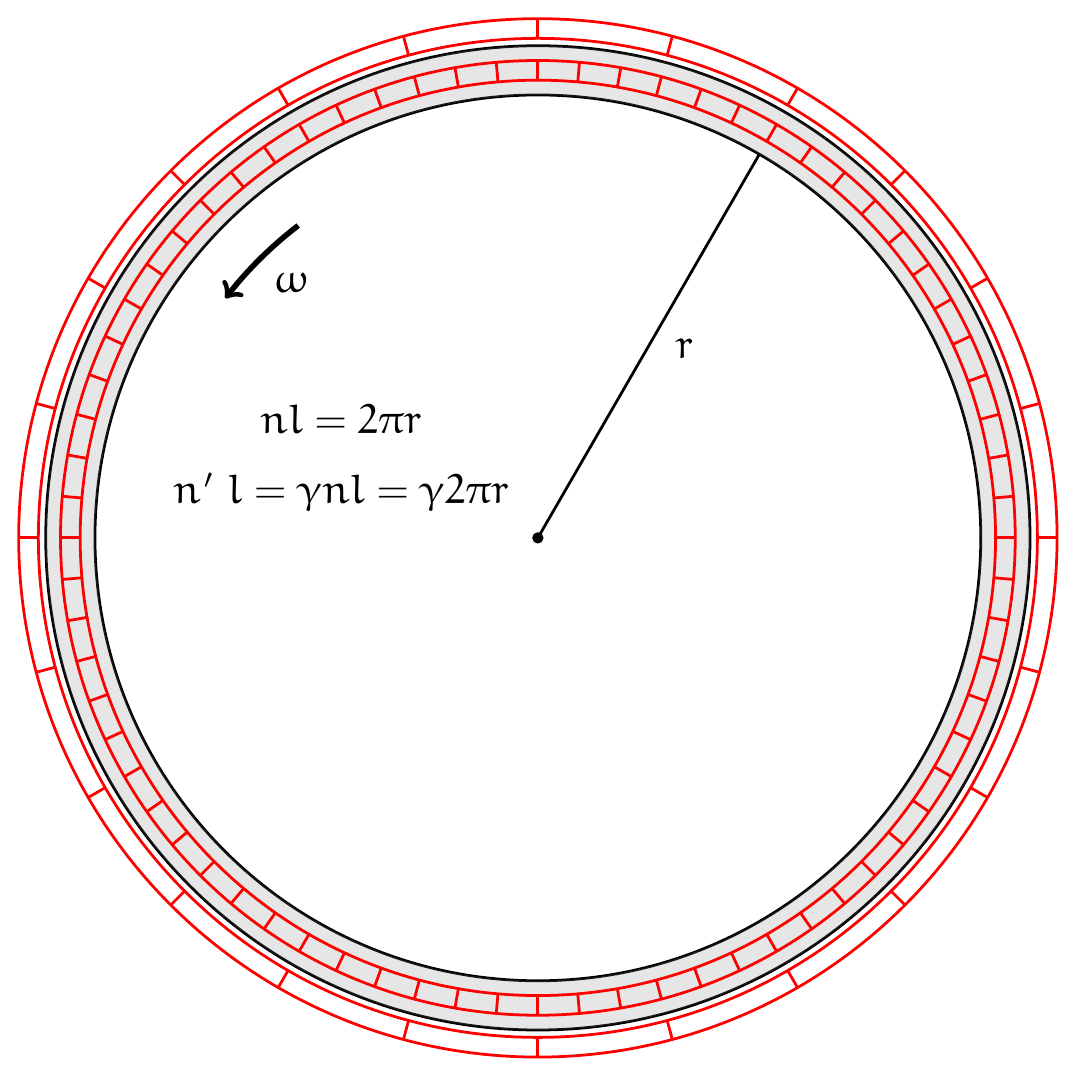}
	\caption{A ring in steady-state rotational motion, shown in grey, which has radius $r$ and circumference $C=2 \pi r$ as measured in the inertial frame. Measuring sticks, which have rest length $l$, are shown in red. Outer measuring sticks are kept at rest in the inertial frame while (inner) measuring sticks kept at the ring are rotating with the ring. Measuring sticks kept at the ring get Lorentz contracted for the inertial frame observer and more number of them are needed to cover the circumference $2 \pi r$. $n$ and $n'$ (with $n'=\gamma n$, where $\gamma= 1/ \sqrt{1-\omega^2 r^2/c^2}$ is the Lorentz factor) are, respectively, the number of outer measuring sticks and the measuring sticks kept at the rotating ring which are needed to cover the circumference in two frames. 
	The thickness of the ring and the measuring sticks can be assumed to be infinitesimally small.}\label{fig1}
\end{figure}

\section{Understanding the paradox}\label{sec2}
In this section, we will consider two scenarios: (a) A ring in steady-state rotational motion, and (b) Transition of a ring from rest to rotational motion. We will show that no paradox arises for the case (a) and the actual paradox posed by Ehrenfest is about the case (b).

\subsection{A ring in steady-state rotational motion: No paradox} \label{sec2a}
It is well known that rotating reference frames have non-Euclidean geometries, as shown first in Refs. \cite{Kaluza, Albert}. The following demonstration of the non-Euclidean geometry introduces ideas that will be useful for a better understanding of the paradox.

Consider a ring which is in steady-state rotational motion with a constant angular velocity $\omega$ relative to an inertial frame $A$ and which has radius $r$ as measured in the inertial frame during the steady-state rotational motion. Its circumference, as measured in the same frame $A$, will be $C=2 \pi r$ as the spatial geometry in an inertial frame is Euclidean. Let us define another frame $K$, the rest frame of the ring, which rotates with constant angular velocity $\omega$ with respect to the inertial frame $A$. We will try to find the ring's radius and circumference as measured by the observers in the rotating frame $K$ and show that the spatial geometry in this frame is non-Euclidean.

In reference frame $A$, measuring sticks at rest are placed around the periphery of the ring. The sticks have an infinitesimally small rest (proper) length $l$, and a large number $n = C/l$ sticks are needed to cover the circumference of the ring in frame $A$ (see Fig. \ref{fig1}). Now, let us put the same measuring sticks with same rest length $l$ on the rotating ring along its circumference so that they are at rest in reference frame $K$. As these measuring sticks have velocity $\omega r$ in reference frame $A$, they will be Lorentz contracted to the length $l'=l / \gamma$ with $\gamma= 1/ \sqrt{1-\omega^2 r^2/c^2}$, when seen from the inertial frame $A$ and a number $n'>n$ of these measuring sticks will be needed to cover the circumference $C$, with $n'=C/l' = \gamma C/l = \gamma n$. But for the rotating frame observers placed at the circumference of the rotating ring, these measuring sticks will have length $l$ as the measuring sticks are at rest with respect to them. So the circumference calculated by the observers in the rotating frame $K$ will be $C'= n' l = \gamma n l = \gamma C$, which is  $\gamma$ times greater than the circumference measured in the inertial frame. The radius of the ring will be same in both the frames as motion is everywhere perpendicular to the radial direction. So, whether we place the measuring sticks along the radial direction in the inertial frame or the rotating frame, the same number of measuring sticks will fit to cover the radius as there will be no change in the length of the measuring sticks. Thus, the ratio of circumference and diameter in the rotating frame $K$ will be $C'/2r =  \gamma C / 2r =  \gamma \pi$ which is greater than $\pi$. This gives rise to the conclusion that the spatial geometry in a rotating frame is non-Euclidean and no paradox arises, in this scenario, for the rotating frame $K$ or the inertial frame $A$ observers.

\subsection{Transition of a ring from rest to rotational motion: Paradox arises} \label{sec2b}
The paradox arises when we try to rotate a ring from rest. In simpler terms the paradox says if a ring of radius $r$ and circumference $C=2 \pi r$, initially at rest in an inertial frame, is rotated to finally give it a constant angular velocity $\omega$, its radius, as measured in the inertial frame, should stay the same as motion is everywhere perpendicular to the radial direction but its circumference, measured in the same inertial frame, should be length contracted by the Lorentz factor $\gamma=1/ \sqrt{1-\omega^2 r^2/c^2}$ as motion is tangential to the circumference, which leads to the contradiction. One might think that there is no contradiction, since the geometry is non-Euclidean as discussed in the Sec. \ref{sec2a}. However, in this case the observations are being made by an inertial frame observer, and the spatial geometry in an inertial frame is Euclidean, so the paradox cannot be resolved via the non-Euclidean geometry in the rotating frame. We will show in Sec. \ref{sec4} that the paradox arises due to the crude application of Lorentz contraction and it disappears when the problem is examined carefully.

\section{Two natural ways of accelerating a rod}\label{sec3}
A ring can be considered as made up of small blocks (or rods) which need to be accelerated tangentially to give the ring some non-zero angular velocity $\omega$. So, let us first analyze two natural ways in which a rod can be accelerated: (a) one for which the rod's length measured in the inertial frame, with respect to which it is being accelerated, stays constant, (b) another for which the rod's rest (proper) length stays constant. This discussion will be very helpful in understanding the resolution of the paradox which is given in later sections.

\subsection{The rod is accelerated by keeping its length, as measured in the inertial frame, constant}\label{sec3a}
It was shown by Singal that the rest (proper) length of a rod increases when all the points of the rod are accelerated simultaneously with respect to the inertial frame in which it was initially at rest \cite{Singal}. In the following consideration, we arrive at the same result by different but much simpler calculations. 

Let us consider a rod of proper length $l$, initially at rest in an inertial frame $A$, which is accelerated, along its length, to the right to give it velocity $v$ with respect to the frame $A$ such that it moves uniformly afterwards. The rod is accelerated in such a way that its length, as measured in frame $A$, remains same as $l$, for which all the points on the rod need to be accelerated simultaneously in frame $A$. The rod is accelerated by applying simultaneous discrete pushes at the ends of the rod (and all the points in between) in frame $A$ until it achieves velocity $v$. It is also maintained that the length of the rod, as measured in frame $A$, stays same as $l$ even after the acceleration program has been switched off and whatever external forces, if any, are needed to maintain that length are provided. The rest frame of the rod after it has finally been accelerated to the velocity $v$ will be an inertial frame, say $B$, which will have relative velocity $v$ with respect to the frame $A$ and the length measured in this frame (frame $B$) will be the rod's proper length. 

After the rod achieves uniform velocity $v$ with respect to frame $A$, we make simultaneous measurements, in frame $A$, at the two ends of the rod so that the time and spatial separation between the measurements are, respectively, $\Delta t_A= 0$ and  $\Delta x_A= l$. Now, by Lorentz transformation, the time separation $\Delta t_B$ and the spatial separation $\Delta x_B$, in frame $B$, between the measurements will be

\begin{equation} \label{eq1}\nonumber
	\begin{split}
	 \Delta t_B & = \gamma \left(\Delta t_A-\frac{v \Delta x_A}{c^2} \right) \\
                & = -\gamma \frac{v l}{c^2}~,\\
	 \Delta x_B & = \gamma \left(\Delta x_A-v \Delta t_A \right) \\
      &=\gamma l~.
	\end{split}
	\end{equation}

Even though the measurements are not simultaneous in frame $B$ (since $\Delta t_B \not= 0$), $\Delta x_B$ gives us the length of the rod in frame $B$ because the rod is at rest in frame $B$. So, the length of the rod in frame $B$, $l_B$, which is also the rest (proper) length of the rod, will be $l_B= \gamma l$. So, we conclude that the proper length of the rod increases when the rod is accelerated by keeping its length in the inertial frame, with respect to which it is being accelerated, constant. This kind of motion is not Born rigid as the proper length of the rod has changed.
Also, for this acceleration program, the two ends of the rod (and all the points in between) are accelerated simultaneously in frame $A$ ($\Delta t_A =0$). But from the above transformations, we find that as soon as the rod achieves some non-zero velocity, however small, pushes at the two ends of the rod do not happen simultaneously in the rod's rest frame - the inertial frame in which the rod is instantaneously at rest - the push at the front end happens earlier than the push at the rear end.

\begin{figure*}[t] 
	\begin{centering}
		\begin{tabular}{cc}
		    \includegraphics[scale=0.7]{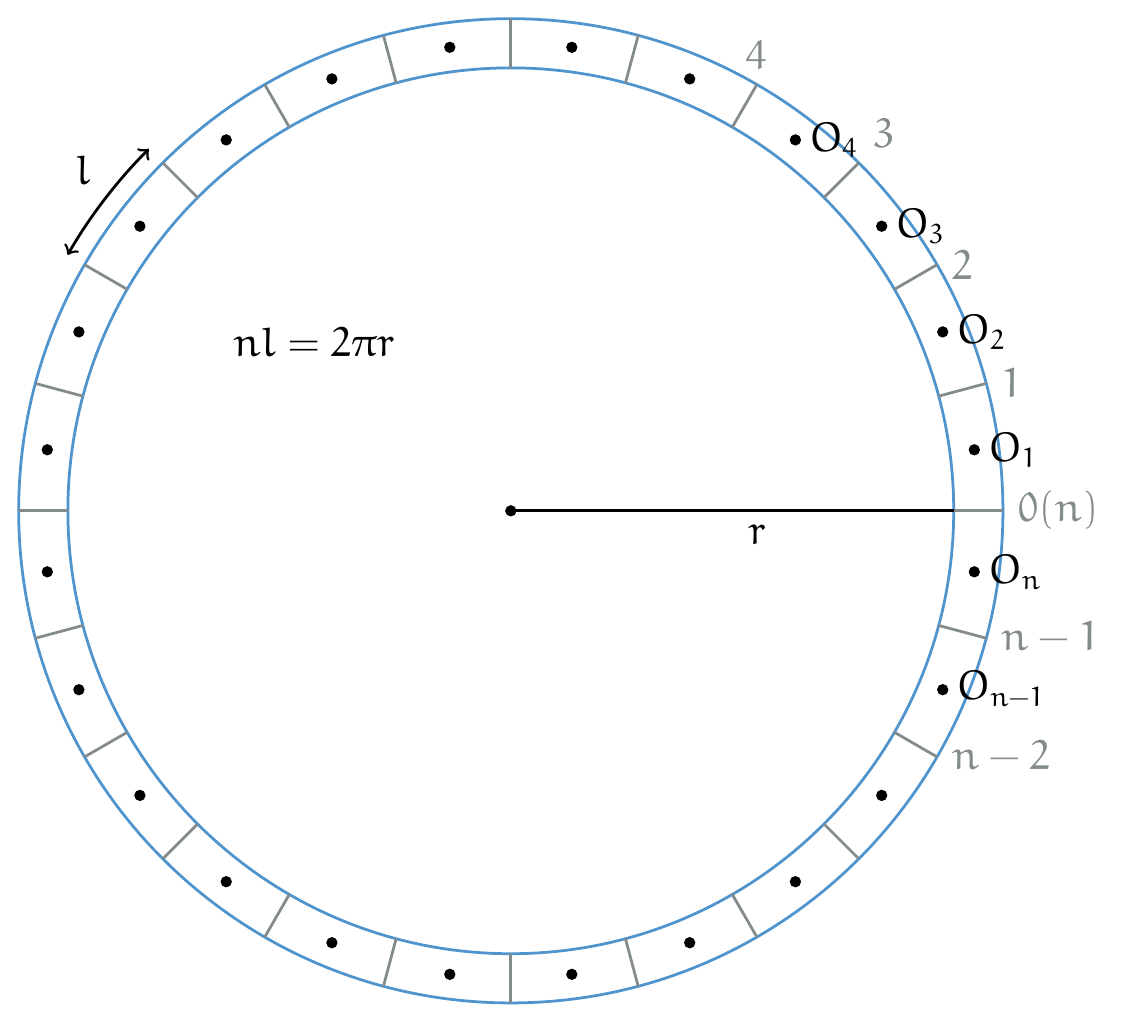}\hspace{2cm}
		    \includegraphics[scale=0.7]{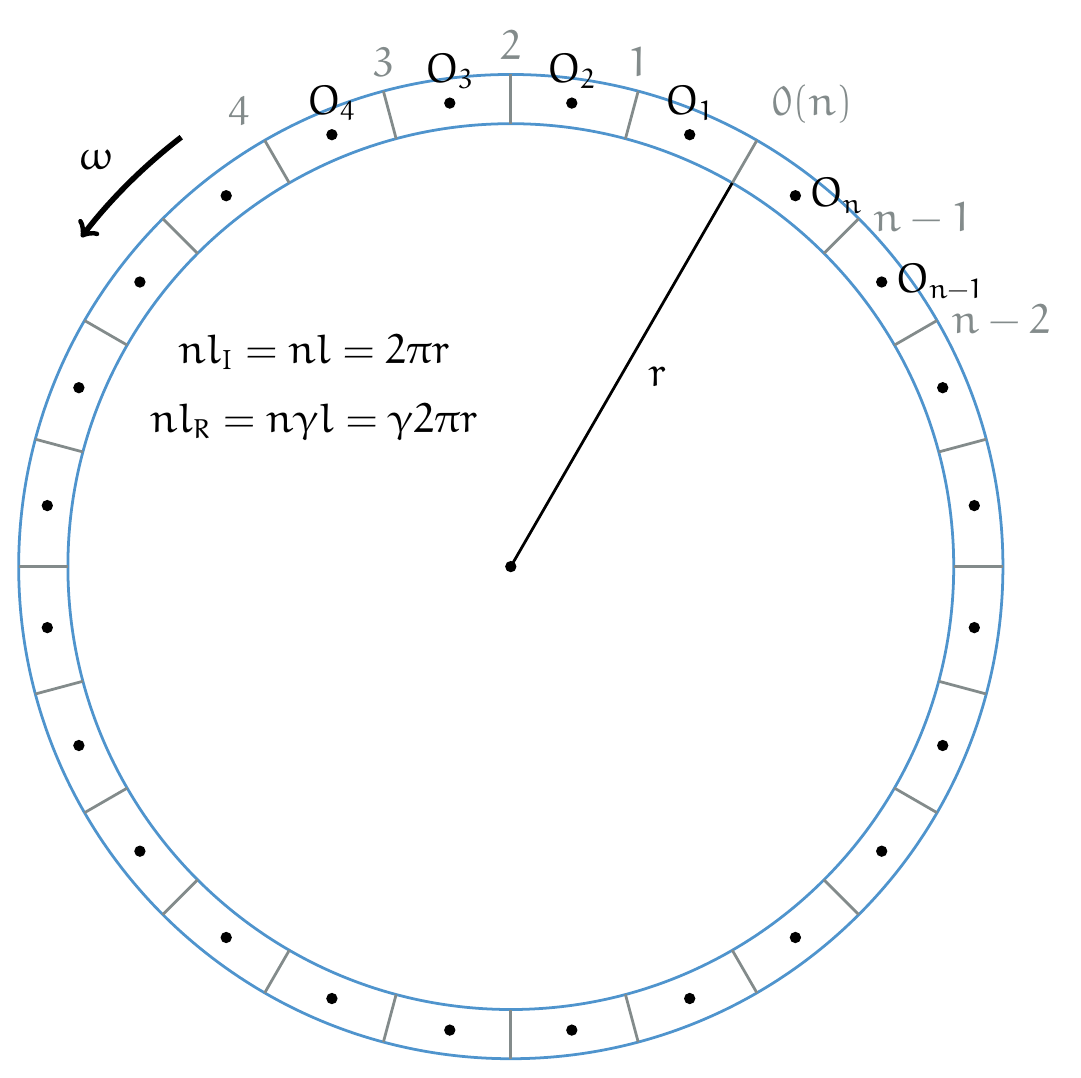}
			\end{tabular}
	\end{centering}
	\caption{(a) (left) A ring, with radius r, at rest in the inertial frame, which has been divided in $n$ number of blocks each having length $l$. $O_i$ ($i=1,2...n$) represents the observer sitting on $i^{\text{th}}$ block while $0, 1...n$ in light grey represent the locations of the ends of the blocks.
	(b) (right) Rotating the ring by keeping the lengths of the blocks, as measured by the inertial frame observer, constant. $l_I$, the length of the (rotating) blocks, as measured by the inertial frame observer, stays same as $l$ due to our chosen acceleration program. $l_R$, the length of the blocks, as measured by the rotating frame observers, i.e., the rest length of the blocks increases by the Lorentz factor $\gamma= 1/ \sqrt{1-\omega^2 r^2/c^2}$.
	}\label{fig2}	
\end{figure*}

\subsection{The rod is accelerated by keeping its rest (proper) length constant}\label{sec3b}

This time the rod is accelerated in a Born rigid way: all the points on the rod are given simultaneous discrete pushes in the instantaneous rest frame of the rod such that the rest length of the rod remains $l$ \cite{Singal}. Finally, when the rod achieves velocity $v$ with respect to the frame $A$ (zero velocity with respect to frame $B$), its length $l_B$, as measured in frame $B$, will be $l$.

In this case also, let us make simultaneous measurements, in frame $A$, at the two ends of the rod after it has been accelerated to the velocity $v$ such that time separation $\Delta t_A$ between the measurements is $\Delta t_A= 0$ while the spatial separation $\Delta x_A$ between the measurements is unknown. Now, by Lorentz transformation, the time separation $\Delta t_B$ and the spatial separation $\Delta x_B$, in frame $B$, between the measurements will be
\begin{equation} \label{eq2}\nonumber
	\begin{split}
	 \Delta t_B & = \gamma \left(\Delta t_A-\frac{v \Delta x_A}{c^2} \right) \\
                & = -\gamma \frac{v \Delta x_A}{c^2}~,\\
	  \Delta x_B & = \gamma \left(\Delta x_A-v \Delta t_A \right) \\
                 & = \gamma \Delta x_A \\
\implies  \Delta x_A & = \frac{\Delta x_B}{\gamma}~.
	\end{split}
	\end{equation}

By earlier arguments, we know that $\Delta x_B$ is the length of the rod in frame $B$ which is the proper length of the rod and is equal to $l$ while $ \Delta x_A$ is the length of the rod as measured in frame $A$. So, the length of the rod in frame $A$, $l_A$, will be $l_A= l/ \gamma$. So, we conclude that the length of the rod in frame $A$ decreases when the rod is accelerated by keeping its proper length constant. Also, for this acceleration program, pushes at the two ends of the rod (and all the points in between) are given simultaneously in the instantaneous rest frame of the rod. But, from the Lorentz transformation, it can be found that as soon as the rod achieves some non-zero velocity, pushes at the two ends of the rod do not happen simultaneously in the inertial frame $A$; the push at the rear end happens earlier than the push at the front end.

\section{Resolution of the paradox}\label{sec4}
Let us now return to the ring of radius $r$ and circumference $C=2 \pi r$, initially at rest in an inertial frame $A$, divided into $n$ blocks, each of infinitesimally small length $l$, such that $ n l = 2 \pi r$. Let there be observers $O_i$ ($i=1,2...n$) in the ring's reference frame, one at each block (see Fig. \ref{fig2}), which will be rotating with the ring when the ring is rotated. For the transition of the ring from rest to rotation, the blocks constituting the ring need to be accelerated tangentially. The resolution of the paradox depends on how the blocks constituting the ring are accelerated: whether they are accelerated by keeping their lengths in the inertial frame $A$ constant, or they are accelerated by keeping their rest (proper) lengths constant. We assume that the blocks are accelerated by giving pushes in the circumferential direction at their ends (remember that these blocks are of infinitesimally small length, so pushes at the two ends are sufficient).

\subsection{Rotating a ring by keeping the lengths of the blocks, as measured in the inertial frame, constant}\label{sec4a}
Let us give simultaneous pushes, according to an observer in the inertial frame $A$, at the ends of the blocks constituting the ring (at points $0,1,2...n-1$) (see Fig. \ref{fig2}) in the circumferential direction until the ring finally achieves constant angular velocity $\omega$. Since the ends of the blocks are being accelerated simultaneously in the inertial frame $A$, the length of each block, as measured by an observer in frame $A$, will remain $l$. And so the circumference, $C_1$, of the rotating ring, as measured in the inertial frame $A$, will be $C_{1}= n l =2 \pi r=C$ which is same as before rotation. The radius, $r_1$, of the rotating ring, as measured in the inertial frame $A$, will also be the same as $r$ due to the motion being everywhere perpendicular to the radial direction. So, there is no paradox for an observer in the inertial frame $A$; for this observer radius as well as the circumference after rotation are same as before rotation. 

Now, let us analyze the situation from the point of view of the observers in the rotating frame $K$ in which the rotating ring is at rest. Due to our chosen acceleration program, the rest (proper) length of the blocks will increase (ref. Sec. \ref{sec3a}) by the Lorentz factor $\gamma$, where $\gamma=1/ \sqrt{1-\omega^2r^2/c^2}$. So, the circumference, $C_1'$, of the rotating ring, as measured by the observers in rotating frame $K$, will be $C_1' = n (\gamma l) = \gamma 2 \pi r= \gamma C$. The radius, $r_1'$, of the rotating ring, as measured in the rotating frame $K$, will be same as $r$ due to the arguments given in Sec. \ref{sec2a}. This implies that the spatial geometry observed by the rotating frame observers will be non-Euclidean, as found in our earlier analysis done for a ring in steady-state rotational motion. So, no paradox arises even for the rotating frame observers, as long as we remember that their geometry is non-Euclidean. 

Even if the ring remains intact, the notion of Born rigidity is not maintained during the transition from rest to rotational motion as the rest length of the periphery, which is the total sum of the rest lengths of the blocks constituting the ring, increases. However, once the ring achieves constant angular velocity $\omega$, the increased rest length of the blocks stays constant and so the motion remains Born rigid. Our analysis is in agreement with Weber's claim that, during the transition from rest to rotational motion, the material of the disk (ring) will physically stretch in the circumferential direction, thus increasing the rest length of the periphery which compensates for its Lorentz contraction \cite{Weber}. 

But what if we try to rotate the ring by keeping the rest lengths of the blocks constant. Should the blocks and consequently the circumference not contract for an observer in the inertial frame? Are we not back to the same paradox? We will analyze this situation in the next section.

\begin{figure} 
		   \includegraphics[scale=0.7]{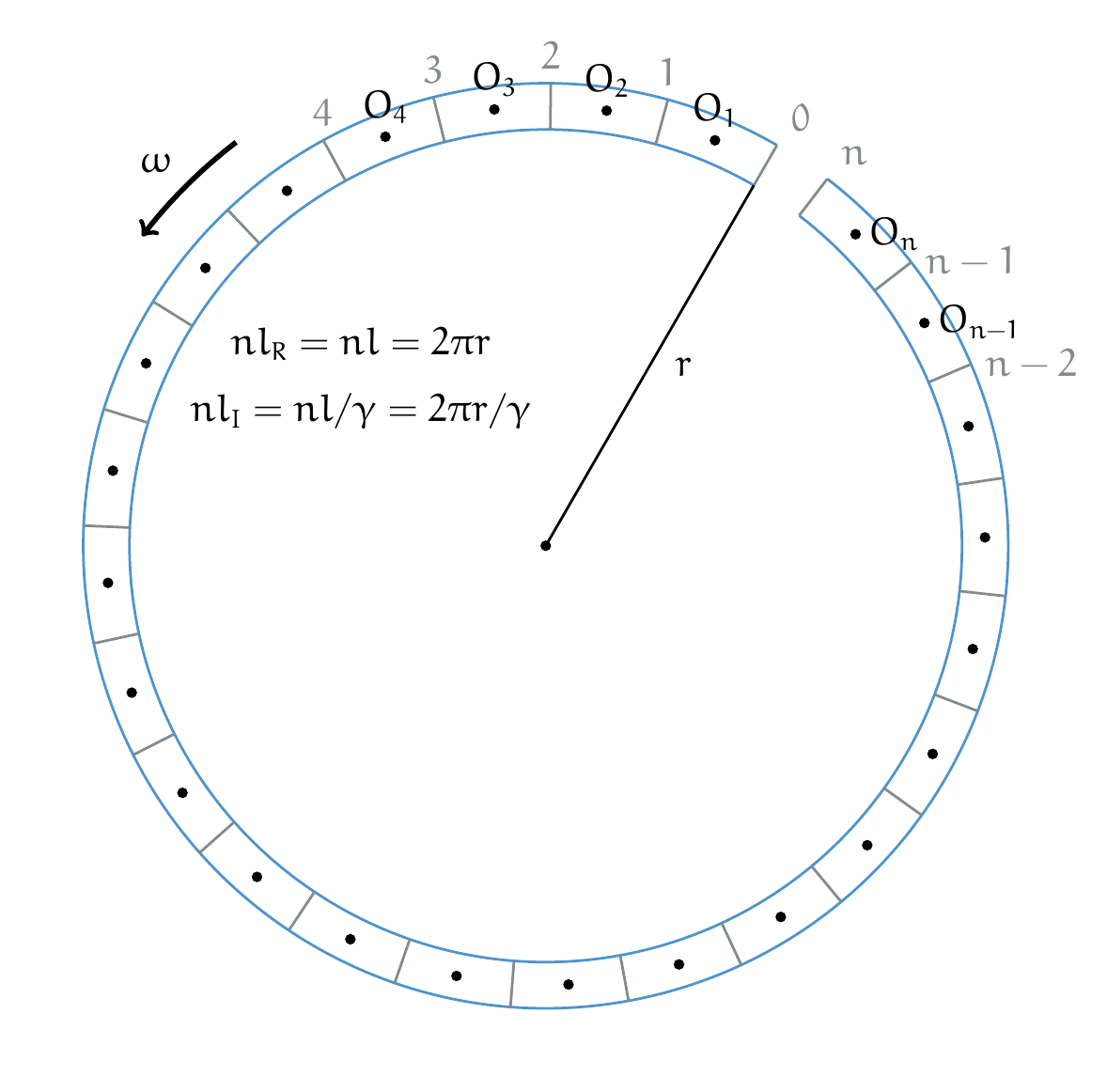}
	\caption{Rotating a ring by keeping the rest lengths of the blocks constant. In this case, $l_R$, the length of the blocks, as measured by the rotating frame observers, i.e., the rest length of the blocks remains the same $l$, while $l_I$, the length of the (rotating) blocks, as measured by the inertial frame observer, gets contracted by the Lorentz factor $\gamma= 1/ \sqrt{1-\omega^2 r^2/c^2}$. Due to our chosen acceleration program the ring tears at the location $0(n)$ and an empty space gap of length $C-C/ \gamma$, in inertial frame, is formed between points $0$ and $n$.   }\label{fig3}
\end{figure}

\subsection{Rotating a ring by keeping the rest (proper) lengths of the blocks constant}\label{sec4b}
As will be shown in this section, it is not possible to rotate a ring from rest, without tearing it, such that the rest lengths of the blocks constituting the ring remain constant. We were able to accelerate a rod this way because for a moving rod there exists an inertial frame in which the rod is instantaneously at rest. But for a rotating ring there is no single inertial frame in which all the blocks constituting the ring will be instantaneously at rest; observers sitting on different blocks belong to different (inertial) rest frames. Let us apply the following acceleration program that Gr{\o}n used to show that the transition of a ring from rest to rotation in Born rigid way is a kinematical impossibility \cite{Gron1975}. The acceleration program keeps the rest length of the blocks constant while also trying its best (but eventually not succeeding) to keep the ring intact. Points $0$ and $1$ are accelerated simultaneously, by applying simultaneous pushes, in the rest frame of the observer $O_1$ such that the rest length of the block $1$ remains fixed (see Fig. \ref{fig3}). Point $2$ is accelerated simultaneous with point $1$ in the rest frame of the observer $O_2$ such that the rest length of block $2$ remains fixed. Point $3$ is accelerated simultaneous with point $2$ in the rest frame of the observer $O_3$ such that the rest length of block $3$ remains fixed. Similarly, points $4,5,6.....n-1$ are also accelerated in the same way. Point $n$ is accelerated simultaneous with point $n-1$ in the rest frame of the observer $O_n$ such that the rest length of block $n$ remains fixed. This process is repeated  until the ring finally achieves constant angular velocity $\omega$. Note that the points $0$ and $n$ coincide when the ring is not rotating.

Now, let us examine this from the point of view of an observer in the inertial frame $A$. For this observer, as soon as the blocks constituting the ring achieve non-zero velocities, however small they might be, pushes at the ends of any block do not happen simultaneously. As we saw in Sec. \ref{sec3b}, for the inertial frame observer, the push at point $1$ happens later than the push at point $0$; the push at point $2$ happens later than the push at point $1$; and so on. And finally, the push at point $n$ happens later than the push at point $n-1$. So the order of pushes in time, from the perspective of the inertial frame observer, is: First at point $0$, then at point $1$, then at point $2$.....then at point $n-1$, then at point $n$. Initially, the points $0$ and $n$ were infinitely close, but as the push at point $0$ happens earlier than the push at point $n$, the ring tears at this location and an empty space gap is formed between points $0$ and $n$. As the ring tears, the notion of Born rigidity is not maintained, in agreement with the statement of Gr{\o}n ``that a transition of the disk from rest to rotational motion, while the circumference satisfies Born's definition of rigidity, is a kinematical impossibility'' \cite{Gron1977}.

After the ring has achieved constant angular velocity $\omega$, let us calculate the radius and the length of the periphery of the ring in the inertial frame $A$ as well as the co-rotating frame $K$. Since the rest length of the blocks is fixed, the length of any block, as measured by an observer in the inertial frame $A$, will decrease by the Lorentz factor $\gamma$. Therefore, the total length of the material ring, as measured by an observer in the inertial frame $A$, will be $C_2=n l / \gamma= C/\gamma$ and the length of the empty space gap, as measured in the same frame $A$, will be $C-C/\gamma$. The radius, $r_2$, of the rotating ring, as measured in frame $A$, will be the same as $r$ as the motion is perpendicular to the radial direction. For the observers in the rotating frame $K$, since the rest length of the blocks stays the same as $l$, the length of the material ring will be $C_2'= n l=C$. As the length of the total circumference, including the empty space gap, in the rotating frame $K$ should be $\gamma C$, the length of the empty space gap in frame $K$ will be $\gamma C - C$. The radius, $r_2'$, of the rotating ring, as measured in frame $K$, will also be the same as $r$ due to the arguments given in Sec. \ref{sec2a}. As the ring tears for this kind of acceleration program, no paradox arises in the inertial frame $A$ or the rotating frame $K$. Therefore, we conclude that the paradox, as stated by Ehrenfest, arose not due to any inconsistency in relativity theory but due to our incorrect assumption that a ring can be brought from rest to rotational motion in a Born rigid way. This is equivalent to two simultaneous assumptions: 
\begin{enumerate}
\item The rest length of the periphery of the rotating ring, which is defined as the summation of the lengths of the infinitesimally small blocks as measured in their respective instantaneous rest frames, remains the same as before rotation and thus leads to the Lorentz contraction of the periphery in the inertial frame. 
\item The ring remains intact i.e., does not tear at any location during the transition from rest to rotational motion. 
\end{enumerate}
We saw that the two assumptions cannot hold simultaneously. If one of the assumptions holds, the other has to be dropped. This analysis is in agreement with Stepanov who showed that when the angular velocity of the ring, $\omega$, acquires time-dependence, the metric tensor for spatial distance in the rotating frame, $\gamma_{ij}$, also becomes evolving in time \cite{Stepanov2018}. So the coordinate system attached with the ring, during the transition of the ring from rest to rotational motion, is no longer locally rigid - i.e., no longer rigid according to Born.

\section*{ACKNOWLEDGMENTS} 
The author expresses gratitude to Professor Sushant G. Ghosh, his Ph.D. advisor, for granting him the freedom to tackle the problem independently. Professor Tabish Qureshi's guidance throughout the submission process is greatly appreciated. The author acknowledges the financial support from CSIR through SRF, award no. 09/466(0203)/2018-EMR-I. Special thanks are extended to the anonymous reviewers and editors for their dedicated time and efforts in enhancing the manuscript.

\section*{AUTHOR DECLARATIONS}
\subsection*{Conflict of Interest}
The author has no conflicts to disclose.

\end{document}